\documentclass[conference, 11pt,letter]{IEEEtran}
\IEEEoverridecommandlockouts
\usepackage{romannum}
\usepackage{amsmath}
\usepackage{graphicx}
\usepackage{color}
\usepackage{multirow}
\usepackage{amssymb,bm}
\usepackage{algpseudocode}
\usepackage{algorithm}
\usepackage{array}
\usepackage{enumerate}
\usepackage{float}
\usepackage{subfigure}
\usepackage{amsthm}
\usepackage{thmtools}
\usepackage[utf8]{inputenc}
\usepackage[T1]{fontenc}
\usepackage{url}
\usepackage{ifthen}
\usepackage{cite}
\usepackage{mathtools}
\usepackage{bbm}
\usepackage{braket}

\declaretheoremstyle[%
spaceabove=6pt,%
spacebelow=6pt,%
headfont=\normalfont\itshape,%
notefont=\normalfont\itshape,
qed=\qedsymbol,%
headpunct={:},
bodyfont=\normalfont,%
]{mystyle}

\declaretheoremstyle[%
spaceabove=6pt,%
spacebelow=6pt,%
headfont=\bfseries,%
bodyfont=\normalfont,%
postheadspace=1em,%
]{mystyle_1}

\makeatletter
\newcounter{phase}[algorithm]
\newlength{\phaserulewidth}

\makeatother

\begin{document}
\bstctlcite{IEEEexample:BSTcontrol}

\pagenumbering{arabic}
\graphicspath{{./resources/}}

\title{Learning to Learn Quantum Turbo Detection}

\author{Bryan Liu\IEEEauthorrefmark{1}\IEEEauthorrefmark{2}\thanks{B. Liu conducted this research during his internship at MERL.}, Toshiaki Koike-Akino\IEEEauthorrefmark{1}, Ye Wang\IEEEauthorrefmark{1}, 
	Kieran Parsons\IEEEauthorrefmark{1} \\

\IEEEauthorblockA{\small\IEEEauthorrefmark{1}Mitsubishi Electric Research	Laboratories (MERL), 201 Broadway, Cambridge, MA 02139, USA.}

\IEEEauthorblockA{\small\IEEEauthorrefmark{2}Electrical Engineering and Telecommunication, University of New South Wales, Sydney, Australia.}
	Email: bryan.liu@unswalumni.com, \{koike, yewang, parsons\}@merl.com
}
\maketitle

\begin{abstract}
    This paper investigates a turbo receiver employing a variational quantum circuit (VQC). 
    The VQC is configured with an ansatz of the quantum approximate optimization algorithm (QAOA). 
    We propose a `learning to learn' (L2L) framework to optimize the turbo VQC decoder such that high fidelity soft-decision output is generated. 
    Besides demonstrating the proposed algorithm's computational complexity, we show that the L2L VQC turbo decoder can achieve an excellent performance close to the optimal maximum-likelihood performance in a multiple-input multiple-output system.  
\end{abstract}


\section{Introduction}
Classical trellis-based codes with the Bahl--Cocke--Jelinek--Raviv (BCJR)~\cite{BCJR} algorithm or low-parity parity-check (LDPC) codes with the belief-propagation algorithm~\cite{LDPC} can achieve or closely approach the maximum a posterior (MAP) decoding performance.
Besides providing an outstanding decoding performance, these soft-decision decoding algorithms can be applied to turbo equalization systems to effectively exchange the extrinsic information between the data detector and channel decoder.
For instance, in a coded multiple-input multiple-output (MIMO) system, a turbo detector consists of a detector with linear and non-linear components and a channel decoder~\cite{wang_poor, oamp_turbo} to iteratively recover the coded linear system.
However, an optimal decoder requires a computational complexity that exponentially increases with the codeword length when a classical computer is used. 
As a result, MAP decoding algorithms introduce a large computational delay.

As an alternative computing paradigm, quantum processors are envisioned to provide the fifth industry revolution. 
Very recently, IBM has successfully manufactured $127$-qubit processors, which are now available in public cloud computing.
With the rapid growth of development on quantum physics and quantum computing, the impracticality of MAP or ML decoding algorithms might be redefined. 
Examples such as Shor's~\cite{Shors} and Grover's~\cite{Grovers} algorithms prove that quantum computing can speedup tasks on integer factorization and unstructured search compared to classical computing. 
Moreover, modern quantum algorithm such as Quantum Approximate Optimization Algorithm (QAOA)~\cite{qaoa} can solve the combinatorial optimization problems by generating a variational quantum circuit (VQC), where the parameters for quantum state preparation are optimized according to the cost and mixer Hamiltonians. 
\cite{qaoa_detection} and~\cite{qaoa_decoding}  have studied QAOA for channel detection and decoding, respectively, to achieve an approximate ML performance with a quantum processor. 
However, the practicality of a quantum processor is still questionable in the following two aspects. 
First, the original QAOA returns a hard-decision ML solution, it is not desirable for modern receiver designs which require each processor to return soft-decision for a precise estimation. 
Second, QAOA relies on a classical meta-heuristic optimization to tune the variational parameters, where
the iterative processing, e.g., stochastic gradient descent process, between a classical computer and a quantum processor might cause an extreme latency in data recovery.

In this paper, we investigate a turbo equalization system with a VQC decoder, where the parameters are directly found by a `learning to learn' (L2L) method with a small number of optimization steps~\cite{l2l}.
Specifically, we consider a coded MIMO system, where the orthogonal approximate message passing (OAMP) algorithm~\cite{oamp, oamp_turbo} serves as the channel detector and the VQC serves as the channel decoder. 
Moreover, the soft-information of each qubit's measurement is converted to the log-likelihood ratio (LLR) to be directly fed back to the detector. 
We call the algorithm as VQC for turbo detection (VQC-TD).
The main contributions of this paper are summarized as follows:
\begin{itemize}
    \item We construct a VQC for channel decoding based on the circuit structure of QAOA ansatz in a coded linear vector system.
    \item We propose an L2L framework to accelerate the process of obtaining the quantum variational parameters for the cost and mixer Hamiltonians.
    \item We compare the bit-error rate (BER) performance between a classical neural network based decoder and VQC-TD. 
    \item We show that our VQC-TD can achieve close to the ML performance.
\end{itemize}

\section{Preliminaries}
\subsection{Channel Model}
In this paper, we assume a linear channel model,
\begin{equation}
    \mathbf{y} = \mathbf{A}\mathbf{x} + \mathbf{w},
\end{equation}
using binary-phase shift-keying (BPSK) modulation with $\mathbf{x} \in \{+1, -1\}^N$ and $\mathbf{A} = \mathbf{U}\mathbf{\Lambda} \mathbf{V}^\text{H}$, where $\mathbf{U} \in \mathbb{R}^{M\times M}$ and $\mathbf{V}^\text{H} \in \mathbb{R}^{N\times N}$ are Haar-distributed matrices. 
$\mathbf{\Lambda}$ contains the singular values for the singular-decomposition form of channel matrix $\mathbf{A}$. 
$\mathbf{w} \in \mathbb{R}^{M}$ represents the additive white Gaussian noise (AWGN) with a zero-mean and variance of $\sigma^2$.
On top of this linear system, we assume that $x_i = (-1)^{v_i}$ for $i \in \{1, 2, \ldots, N\}$. 
$\mathbf{v} = \mathbf{G}\mathbf{u} \pmod  2$ represents a valid codeword in the codebook $\mathcal{C}$, which has a codeword length of $N$, information bits length of $K$ and number of syndromes of $S$. 
$\mathbf{G}$ represents the generator matrix of the channel codes. 
Therefore, besides performing a linear recovery process to estimate the transmit symbols $\mathbf{x}$, we expect to decode the information bits $\mathbf{u}$.

\subsection{Quantum Approximate Optimization Algorithm}

QAOA was proposed in \cite{qaoa} to solve combinatorial optimization problems. 
According to the objective function of the problem to be solved, QAOA constructs a VQC, where the deeper the circuit, the better approximation can be observed. 

In QAOA, there are two types of Hamiltonians that are concatenated and repeated several times for quantum state-evolution. 
One is defined as cost Hamiltonian $H_C$, and the other is defined as mixer Hamiltonian $H_M$. 
Cost Hamiltonian $H_C$ is determined based on the objective function to be optimized and the mixer Hamiltonian $H_M$ adjusts the state in each layer so that the VQC is time evolved with the increasing depth of layers. 
For a quantum circuit constructed from unitary operation, $e^{i\alpha H}$, where $\alpha$ is a scalar and $H$ is a Hamiltonian, 
the cost Hamiltonian $H_C$ and mixer Hamiltonian $H_M$ introduce a variational parameter $\gamma_d$ and $\beta_d$, respectively to the $d$th layer of the VQC.
Define $D$ as the depth of VQC generated by QAOA.
The unitary operation of the quantum circuit depends on the parameters $\bm{\gamma} = [\gamma_1, \gamma_2, \ldots, \gamma_D]$ and $\bm{\beta} = [\beta_1, \beta_2, \ldots, \beta_D]$. 
We use
\begin{equation}
    U(\bm{\gamma}, \bm{\beta}) = e^{i\beta_DH_M}e^{i\gamma_DH_C} \cdots e^{i\beta_1H_M}e^{i\gamma_1H_C},
\end{equation}
to represent the unitary function of a QAOA circuit with depth $D$.
Suppose the QAOA circuit has an initial state $\ket{v}$.
The expectation on the cost Hamiltonian can be computed by:
\begin{equation}
    \mathcal{L} = \bra{v}U^\dag(\bm{\gamma}, \bm{\beta}) H_CU(\bm{\gamma}, \bm{\beta})\ket{v},
\end{equation}
where $U^\dag$ represents the Hermitian adjoint of the unitary operator.
As a result, the approximate solution of QAOA is found by optimizing the objective function $\mathcal{L}$, which depends on $\bm{\gamma}$ and $\bm{\beta}$.

\subsection{Turbo Detection with OAMP}
The extrinsic-message-aided OAMP (EMA-OAMP) algorithm extends the concept of OAMP algorithm for turbo linear recovery problem \cite{oamp_turbo, oamp}. 
The non-linear detection step combines both the estimate from the linear detection and the extrinsic message from the decoder to explore the information from the transmit symbols and encoded codeword.
Specifically, the EMA-OAMP algorithm follows the iterative processes of:
\begin{align}
    &\text{(Linear Detector):}\ &\mathbf{l}^{t+1} &= g^\text{OAMP}(\mathbf{p}^t), \label{LD}\\ 
    &\text{(Decoding):}\ & \mathbf{q}^{t+1} &= \mathcal{D}(\mathbf{l}^{t+1}), \label{Decoder} \\ 
    &\text{(Non-Linear Detector):}\ & \mathbf{p}^{t+1} &= \eta^\text{OAMP}(\mathbf{l}^{t+1}, \mathbf{q}^{t+1}). \label{NLD}
\end{align}
Here, $\mathbf{p}^0 = \mathbf{0}$ is initialized as the non-linear detection estimate. 
$\mathbf{l}^{t+1}$ and $\mathbf{q}^{t+1}$ refer to the linear estimate of $\mathbf{x}$ and decoder output, respectively.
The linear detector in \eqref{LD} follows the original OAMP algorithm\cite{oamp} and its estimate can be considered as a sequence of transmitted symbols that are corrupted by AWGN. 
Therefore, \eqref{Decoder} can be any channel decoder's extrinsic information with an input channel of AWGN.
Note that \eqref{NLD} differs from the non-linear estimator in the original OAMP algorithm\cite{oamp}. 
In particular, \eqref{NLD} has an additional input of extrinsic information from the channel decoder.
Suppose a minimum mean-square error estimator (MMSE) is used in \eqref{NLD}, 
where the original estimator without extrinsic information is
defined as $\eta^\text{MMSE}(l_i^t) = \text{E}\{x_i | l_i^t\}$. 
With the extrinsic information from the decoder, the MMSE estimator is modified to
\begin{equation}\label{mmse}
    \eta^\text{MMSE}(l_i^t, q_i^t) = \sum_{s'}  \frac{r_{s'} q_i^t(r_{s'})P(l_i^t | x_i=r_{s'})}{\sum_{s} q_i^t(r_s)P(l_i^t | x_i=r_s)}.
\end{equation}
In \eqref{mmse}, $s$ indicates the index of the set of modulation symbols and $r_s$ represents the corresponding symbol that is used in the linear system. 
$q_i^t(r_s) = P(x_i=r_s | \mathbf{l}^t_{\sim i})$ represents the extrinsic information from the channel decoder for the modulation symbol $r_s$. 
Refer \cite{oamp_turbo} for detailed derivations of the EMA-OAMP algorithm.

\section{Turbo Quantum Detection: VQC-TD}
Achieving MAP decoding performance with a classical computer is generally difficult for a practical receiver design. 
With the development of quantum processors, those MAP decoding algorithms may be achieved faster with lower power consumption.
In this paper, we investigate a VQC generated by QAOA and the VQC serves as the source of extrinsic information generator in \eqref{Decoder}.
Moreover, combining the syndrome information, we propose a learning to learn structure for the ease of finding the trainable parameters in VQC. As a results, traditional optimization scheme, stochastic gradient descent, is not required to find the trainable parameters in a VQC.

\subsection{Cost Hamiltonian for VQC-TD}
To begin with, we first introduce the method for finding a VQC for channel decoding. Following the idea in \cite{qaoa_decoding}, a cost Hamiltonian can be constructed from the ML function of $\underset{\mathbf{u}}{\mathrm{argmin}} \sum_{i=1}^N (l_i^t - (-1)^{v_i})^2$ for $\mathbf{v}=\mathbf{G}\mathbf{u}$. 
By expanding the ML function, we obtain the cost Hamiltonian as:
\begin{equation}
    H_C = -\sum_{j=1}^Nl_j^t\prod_{\{i|G_{ji = 1}\}}Z_i,
\end{equation}
where $l_j^t = x_j + z_j$ is considered as the BPSK symbol corrupted by AWGN with $z_j \sim \mathcal{N}(0, \sigma_z^2)$.
We use $Z_i$ to denote the Pauli-Z gate observable of the $i$th qubit. 
Then, the cost Hamiltonian $H_C$ is represented by $e^{i\gamma_dH_C}$ for each layer in the VQC. 
Following \cite{qaoa}, the mixer Hamiltonian $H_M$ is represented by a rotation-X gate for each layer.
The deeper the VQC, the more parameters that can be equipped for the quantum state preparation and the better approximate performance is expected to be obtained after tuning the parameters, $\bm{\gamma}$ and $\bm{\beta}$~\cite{qaoa}.

As a proof-of-concept, we propose a framework that uses VQC as a source to provide the soft-decision information from a decoder.
Note that \cite{qaoa_decoding} and \cite{qaoa_detection} consider the channel decoding and detection separately. 
A combined cost Hamiltonian is straightforward to evaluate the joint effect of both the generator matrix $\mathbf{G}$ and channel matrix $\mathbf{A}$. 
In this paper, we instead tackle the issue by turbo detection framework.

A classical optimizer such as stochastic gradient descent algorithm can be used to adjust the parameters but could lead to a large number of optimization steps. 
\cite{l2l} introduces a learning to learn (L2L) method to find the pattern of optimal $\bm{\gamma}$ and $\bm{\beta}$ by employing a classical recurrent neural network. 
The L2L method largely reduces the required number of optimization iterations compared to a classical optimizer.
However, the framework proposed in \cite{l2l} is lacking in exploration of code structure.
Therefore, in this paper, we adopt the L2L method under the scenario of channel decoding.

\subsection{Learning to learn (L2L) for VQC-TD}
Instead of using traditional stochastic gradient descent method to gradually tune the trainable parameters in VQC, which requires a large number of information exchanges between a classical computer and a quantum processor, we propose a learning to learn scheme to find the trainable parameters in a small number of information exchanges.
\subsubsection{Input parameters}
The L2L method was proposed to accelerate the optimization steps while finding the trainable parameters in the VQC~\cite{l2l}. 
In specific, for the scenario of learning the parameters to be used in QAOA, the input of a classical neural network is the initialization of $\bm{\gamma}$, $\bm{\beta}$ and the expectation over the QAOA's cost Hamiltonian. 
The output of the learning to learn method is the updated values of $\bm{\gamma}$ and $\bm{\beta}$.

\begin{figure}[t!]
    \centering
    \includegraphics[width=\linewidth]{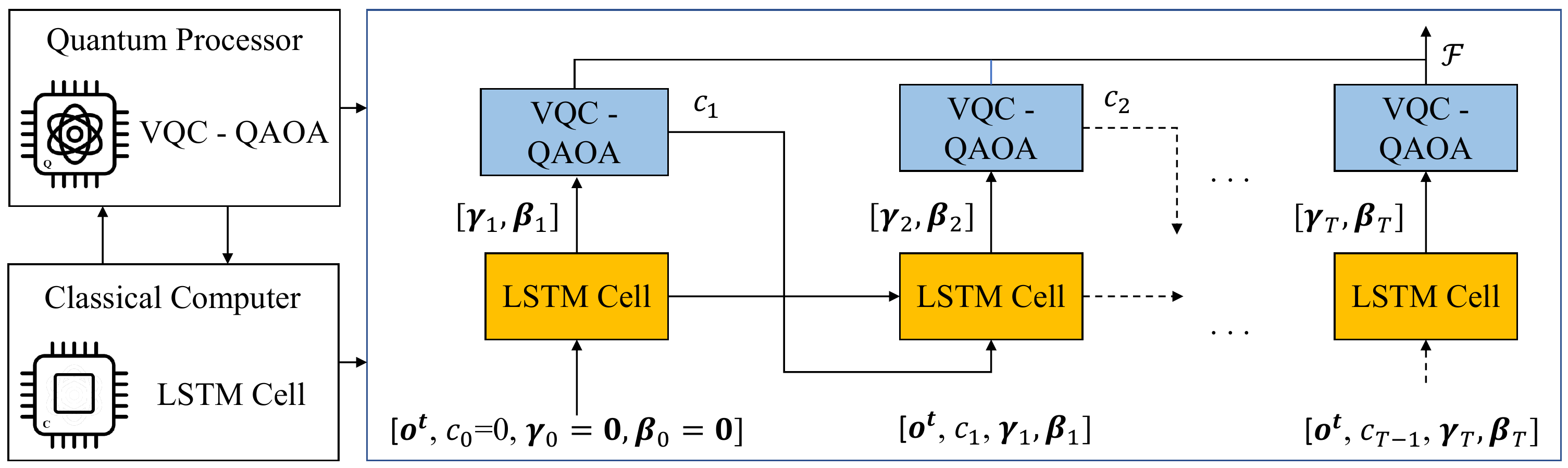}
    \caption{Learning to learn (L2L) framework for VQC-TD.}
    \label{L2L_QAOA}
\end{figure}

In \cite{l2l}, a series of epochs is performed to optimize $\bm{\gamma}$ and $\bm{\beta}$ through a long short-term memory (LSTM) recurrent neural network. 
To accelerate the training process of a neural network, inspired by the syndrome-based decoding algorithm proposed in \cite{syndrome_based}, we use additionally the syndromes and absolute of channel information as the L2L input.
As a result, the LSTM cell is expected to return an accurate prediction on $\bm{\gamma}$ and $\bm{\beta}$ in one epoch.
Fig.~\ref{L2L_QAOA} depicts the overall process of L2L framework to optimize VQC.

\begin{figure}[t!]
    \centering
    \includegraphics[width=\linewidth]{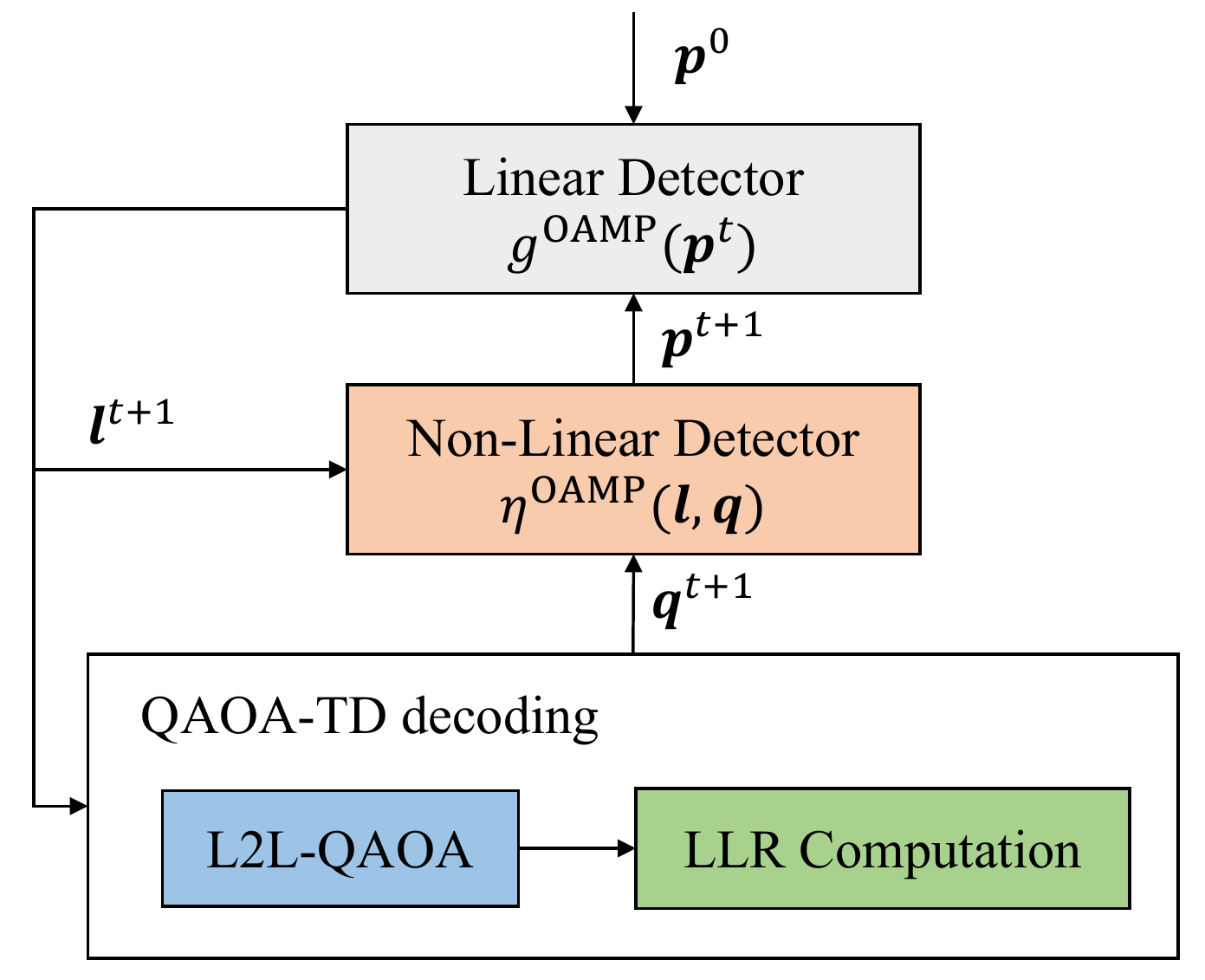}
    \caption{VQC-TD with LLR tuning.}
    \label{VQC-TD}
\end{figure}

Let the syndromes at the $t$th iteration of EMA-OAMP be $\mathbf{s}^t = \mathbf{H}f_b(\mathbf{l}^t) \pmod 2$, where $f_b(l_i^t) = \frac{1}{2}(1-\text{sign}(l_i^t))$ indicates a hard-decision function that maps $l_i^t$ to its binary bit.
During our experiment, we noticed that a weighted syndrome
\begin{equation}
    s_i'^t = \min_{j | H_{ij}= 1}\{|l_j^t|\}
\end{equation}
for $i \in \{1,2,...,K\}$ provides a much faster convergence speed than using the binary syndrome information during the training process. 
Therefore, we use the weighted syndromes as the constant input of the LSTM recurrent neural network.
As shown in Fig.~\ref{L2L_QAOA}, besides the cost Hamiltonian and rotation  parameters $\bm{\gamma}_{t-1}$ and $\bm{\beta}_{t-1}$, we use $\mathbf{o}^t = \{\mathbf{s}'^t, |\mathbf{l}^t|\}$ to denote the additional set of information to be processed by the LSTM cell, where $|\mathbf{l}^t|$ is the element-wise absolute value over the linear estimate $\mathbf{l}^t$.
In total, there are $T$ time steps to update the parameters.

\subsubsection{Loss function}
Since the extrinsic information from the QAOA decoder is expected to be transferred back to the channel detector, we use a binary cross-entropy loss function to train the LSTM cell so that the measurement over the qubits can represent the LLR of each information bit.
Define $\hat{\mathbf{u}}^i \in \mathbb{R}^K$ as the average measurement results of each qubit over a Pauli-Z basis at the $i$th time step for $i \in \{1, 2, \ldots, T\}$.
We consider an exponentially decaying cross-entropy function so that later time step's output of the LSTM cell will contribute more to the loss function.
The final loss function for L2L is thus defined as 
\begin{align}
    \mathcal{F} = 
    -\sum_i^T\sum_j^K \xi^{T-i}\bigg{(}u_j\log\Big{(}\frac{1-\hat{u}_j^i}{2}\Big{)} + \notag \\
    (1-u_j)\log\Big{(}\frac{1+\hat{u}_j^i}{2}\Big{)}\bigg{)},
\end{align}
where $\xi$ is a decay parameter to adjust the contribution of cost at each time step.
Since $\hat{\mathbf{u}}^i$ is measured by a Pauli-Z gate with a value range from $-1$ to $1$, we use $(1-\hat{u}_j^i)/2$ and $(1+\hat{u}_j^i)/2$ to represent the probabilities of measuring $\ket{1}$ and $\ket{0}$ of the $j$th qubit at the $i$th time step, respectively.

\begin{table}[t]
\begin{algorithm}[H]
\normalsize
\caption{VQC-TD Algorithm}
\label{algo}
\textbf{Input}: Linear estimate $\mathbf{l}^t$, mean square error of linear estimate $\tau_t^2$. \\
\textbf{Output}: Extrinsic information $\mathcal{D}_\text{VQC-TD}(\mathbf{l}^t)$
\begin{enumerate}[\bfseries Step 1:]
\addtolength{\itemindent}{6mm}
    \item Find $\mathbf{o}^t$ by computing the syndromes $\mathbf{s}$ and the normalized absolute of channel information $|\mathbf{l}^t|$.
    \item Initialize the QAOA cost $c_0$, trainable parameters $\bm{\gamma}$ and $\bm{\beta}$.
    \item Perform $T$ steps of LSTM and VQC.
    \item Measure the LLRs of $K$-qubits by \eqref{L_info}.
    \item Compute the LLRs of $N$ symbols by \eqref{L_code}.
    \item Return the extrinsic information $L_j^\text{ext} = L_j' - {2l_j^t}/{\tau_t^2}$ for $j \in \{1, 2, \ldots, N\}$ back to the non-linear detector.
\end{enumerate}
\end{algorithm}
\end{table}

\subsubsection{Log-Likelihood Ratio Computation}
In the VQC, each qubit represents the decoded information bit. 
Define $L_i$ as the LLR of measuring $\ket{0}$ over $\ket{1}$ of the $i$th information bit, i.e.,
\begin{equation}\label{L_info}
    L_i = \text{log}\Bigg{(}\frac{1 + \hat{u}_j^T}{1 - \hat{u}_j^T}\Bigg{)},
\end{equation}
for $i \in \{1, 2, ..., K\}$.
Then, let $L'_j$ be the LLR of encoded bit for $j \in \{1, 2, ..., N\}$. 
The LLR of the encoded bit is computed by the belief-propagation updating rule of
\begin{equation}\label{L_code}
    L'_j = 2\,\tanh^{-1}\bigg{(}\prod_{\{i|G_{ij}=1\}}\tanh\Big{(}\frac{L_i}{2}\Big{)} \bigg{)},
\end{equation}
for $j \in \{1, 2, ..., N\}$.

VQC-TD firstly trains the L2L LSTM cell to provide an estimate on the LLR of the encoded information bit. 
Then, the information bit's LLR is used to compute the encoded bit's LLR before returning the extrinsic information back to the non-linear OAMP detector. 
The VQC-TD algorithm is summarized in Algorithm~\ref{algo}.

\begin{table}[t]
	\centering
	\caption{Hyper-parameters of training VQC-TD}
	\label{hyperparameters}
		\begin{tabular}{c c}
			\hline 
			Parameters & Values \\
			\hline
			Exponential decay $\xi$   & $0.6$ \\
			Learning rate & $0.008$ \\
			LSTM depth $L_D$ & $3$ \\
			 LSTM time steps $T$ & $15$ (LDPC), $19$ (BKLC) \\
			QAOA depth $D$ & 18 (LDPC), $22$ (BKLC) \\
			Optimizer & Adamax\\
			\hline	
		\end{tabular}
\end{table}

\section{Numerical Results}

\subsection{System Parameters}

We evaluate the BER performance of joint ML detection and decoding, classical syndrome-based decoder, and VQC-TD on recovering the coded MIMO system.
The channel matrix $\mathbf{A}$ is assumed to have a condition number of $1$ and the singular value $\lambda_i$ is normalized to $\sum_{i=1}^M\lambda_i=N$.
The hyper-parameters for training the VQC-TD are given in Table~\ref{hyperparameters}.
We note that L2L needed just about $20$ time steps for finding optimized parameters $\bm{\gamma}$ and $\bm{\beta}$. 
It means that a classical computer interacts $20$ times with the quantum processor to gradually find the parameters $\bm{\gamma}$ and $\bm{\beta}$. 
We consider two types of channel codes: a regular ($2$, $4$) LDPC code, and a best-known linear code (BKLC)\cite{bosma1997magma} for $N = 20$ and $K = 11$.

\subsection{Bit Error Rate Performance}

\begin{figure}[t!]
    \centering
    \includegraphics[width=\linewidth]{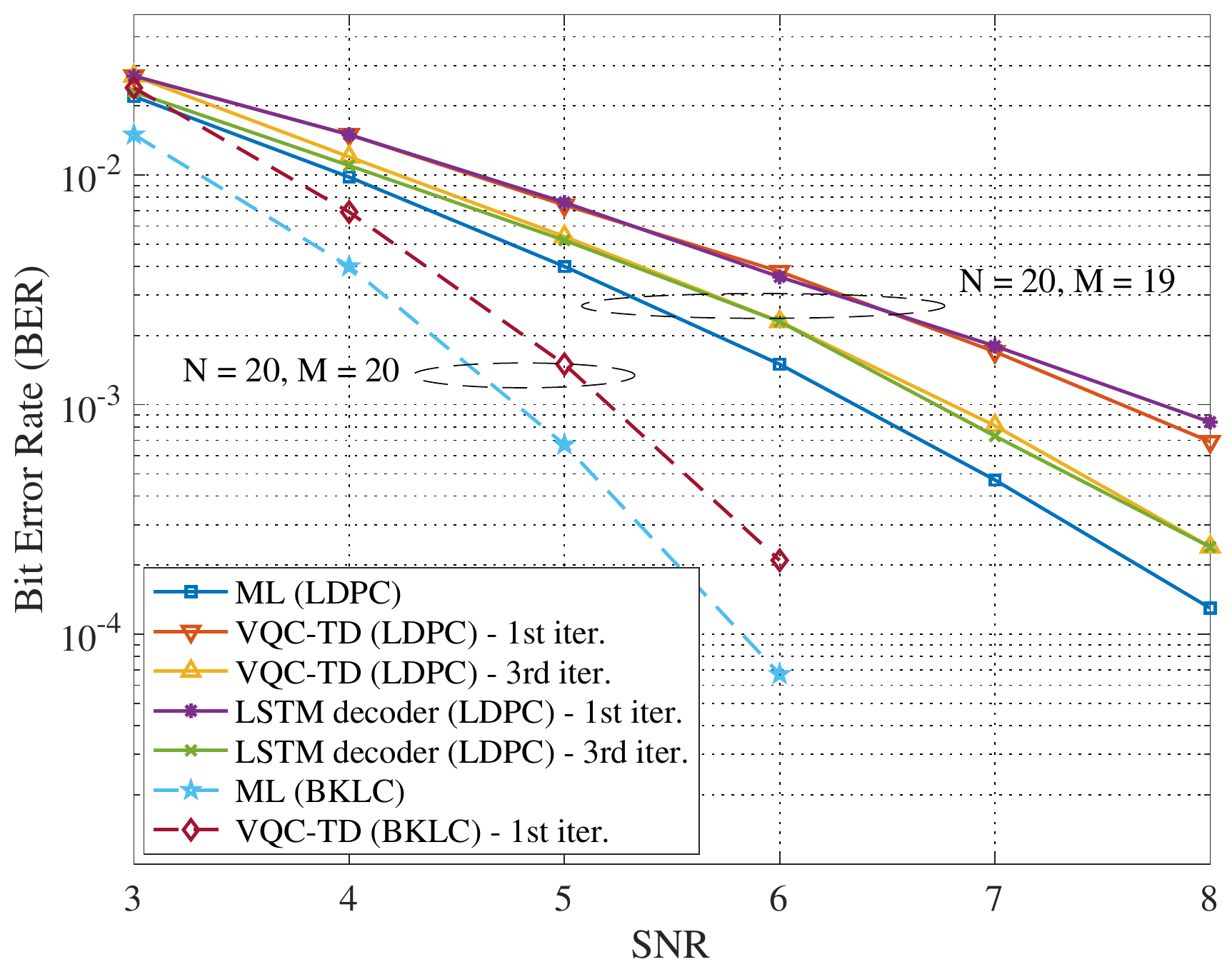}
    \caption{BPSK modulation with $N=20$ with
    LDPC code and  BKLC.}
    \label{BER_N20_M19}
\end{figure}

In Fig.~\ref{BER_N20_M19}, ``ML'' indicates the exact joint ML detection and decoding performance that returns $\mathrm{argmin}_\mathbf{u} \sum_{i=1}^N (y_i^t - r_i)^2$ for $\mathbf{r} = \mathbf{A}(-1)^{\mathbf{G}\mathbf{u} \pmod{2}}$.
``LSTM decoder'' refers to the syndrome-based decoder in \cite{syndrome_based}. 
The LSTM decoder has a depth of $3$, number of time steps of $7$ and a hidden layer size of $150$.
The LSTM decoder is used to generate the extrinsic information from the decoder to the detector.
Then, ML decoding is performed to measure the BER.
From Fig.~\ref{BER_N20_M19}, we can observe that VQC-TD has a close to the ML performance within $0.5$ dB after $3$ iterations of EMA-OAMP. 

Note that the codeword length we consider here is relatively short, which might break the assumption of OAMP algorithm, where each element in the linear estimate $\mathbf{l}^t$ is independent and identically distributed variable. 
Therefore, we choose $N = M = 20$ for BKLC to decorrelate the elements in the linear estimate $\mathbf{l}^t$ so that the closeness between VQC-TD and ML performances can be better estimated.
Similar to the numerical results we have observed for decoding LDPC code, VQC-TD has a close performance to the ML performance within $0.5$ dB.

\begin{table}[t]
	\centering
	\caption{Computational Complexity of VQC-TD per time step}
	\label{QAOA_TD_complexity}
		\begin{tabular}{c c}
			\hline 
			Processors & Number of parameters\\
			\hline
			LSTM Cell   & $4L_D(8D^2 + 2DS+2DN+4D)+4D^2+2D$ \\
			QAOA VQC & $2D$\\
			\hline	
		\end{tabular}
\end{table}

\subsection{Complexity Analysis}
Here we discuss the computational complexity of our proposed VQC-TD algorithm. 
As shown in Table~\ref{QAOA_TD_complexity}, VQC-TD contains two processors, the LSTM cell for L2L and the VQC for turbo decoding.
For the L2L, the LSTM cell has an output layer of size $2D$ depending on the depth of QAOA.
We let $R_Z$, $R_X$ and $H$ denote the number of rotation-Z, rotation-X and Hadamard gates in the VQC, respectively.
Define $N_c$ as the average column weight of a generator matrix $\mathbf{G}$.
The VQC has a total of $D\Big{(}N_c(R_Z^{\otimes N_c}) + K(R_X + H)\Big{)}$ operational gates, where $R_Z^{\otimes N_c}$ indicates the multi-Z rotations across $N_c$ qubits.
Note that the number of trainable parameters for the operational gates is $2D$.
We can observe that the LSTM cell's complexity increases quadratically with the depth of VQC and the average of column weight in the generator matrix $\mathbf{G}$ contributes most to the number of quantum gates in the VQC.

\section{Conclusion}
In this paper, we propose an L2L-VQC framework that is applicable for turbo detection in a MIMO system. 
The algorithm consists of a quantum circuit and a conventional LSTM cell to evaluate the proper rotation angles in the rotation gates of VQC. 
We show that the proposed VQC-TD algorithm can achieve close to ML performance.
To the best of our knowledge, this paper is the first study of a quantum turbo processor.

\bibliographystyle{IEEEtran}
\bibliography{refs}
\end{document}